# Population Density as an Equalizing (or Misbalancing) Mechanism for Species Coexistence

Rafael D. Guariento[1]*

* Corresponding author: *E-mail address*: rafaguariento@gmail.com (Rafael Dettogni Guariento)

*Present Adresses:*

1 - Universidade Federal do Rio Grande do Norte, Departamento de Engenharia Civil, Rio Grande do Norte, Brazil

**Abstract**

Despite the general acknowledgment of the role of niche and fitness differences in community dynamics, species abundance has been coined as a relevant feature not just regarding niche perspectives, but also according to neutral perspectives. Here we explore a minimum probabilistic stochastic model to evaluate the role of populations relative initial and total abundances on species chances to outcompete each other and their persistence in time (i.e., unstable coexistence). The present results show that taking into account the stochasticity in demographic properties and conservation of individuals in closed communities (zero-sum assumption), population initial abundance can strongly influence species chances to outcompete each other, and also influence the period of coexistence of these species in a particular time interval. Systems carrying capacity can have an important role in species coexistence by exacerbating fitness inequalities and affecting the size of the period of coexistence. This study also shows that populations initial abundances can act as an equalizing mechanism, reducing fitness inequalities, which can favor species coexistence and even make less fitted species to be more likely to outcompete better fitted species, and thus to dominate ecological communities in the absence of niche mechanisms.

**Introduction:**

For many decades, models have been proposed to describe and explain observed patterns in species abundances (Fisher et al. 1943, Preston 1948 1962, MacArthur 1957, 1960 Hubbell 2001, Etienne and Olff 2005), but consensus on a single adequate model has not been reached. Much of ecology is built on the assumption that species differ in their niches. Classic studies have shown that species may differ in their use of multiple-limiting resources (Tilman 1982, Grant 1986), their ability to colonize disturbed sites (Grubb 1977), and their response to temporal fluctuations in the environment (Caceres 1997). Current knowledge claims that niche differences cause species to limit their own populations more than they limit others (stabilizing effect), supporting coexistence (Chesson 2000). One good example for the effect of stabilizing process is the role of population abundance (i.e., relative frequency) in reducing population average per capita growth rates (Adler et al. 2007), making a species to suppress itself faster relative to a competing species.

Neutral theory directly challenges the niche paradigm by proposing that high diversity of many natural communities can be achieved assuming species equivalences. However, it does not imply that the world in in fact neutral, instead it does suggest progress can be made incorporating neutral perspective (Rosindell et al. 2012). The idea that neutral processes regulate ecological communities (Bell 2001, Hubbell 2001) has had a profound effect on ecology, but this remains contentious (Gaston & Chown 2005). Data-driven studies generally refute some aspect of neutral patterns or processes (McGill 2003; Gilbert & Lechowicz 2004; Turnbull et al. 2005; Wootton 2005), denying that neutral dynamics can produce observable ecological patterns. Similarly, completely niche-based explanations have failed to adequately explain extant community patterns (Chave 2004, Holyoak and Loreau 2006). Consequently, a number of studies have attempted to reconcile neutral and niche dynamics (Chave 2004, Tilman 2004, Gravel et al. 2006, Holyoak and Loreau 2006, Adler et al. 2007). The divergence and reconciliation between neutral and niche dynamics can be described viewing coexistence mechanisms as either equalizing or stabilizing (Chave 2004, Adler et al. 2007). Stabilizing coexistence describes species differences that result in reduced niche overlap, thus minimizing the impact of fitness inequalities on competitive interactions. Equalizing mechanisms promote similarities in species responses (i.e. fitness equivalency), reduces the rate of competitive exclusion, and promote coexistence from weak stabilizing mechanisms. Fitness differences, by contrast, are differences that drive competitive dominance. These include differences between species that, on average, favors one competitor over the other. Examples include differences in the species' abilities in reducing common limiting resources (Tilman 1988), differences in predator susceptibility (Mitchell & Power 2003) or variation in the number of offspring produced per parent (Stibor 1992). Differences between species in these traits are not niche differences – the advantage does not depend on the relative abundances of the species involved (Adler et al. 2007). Rather than stabilizing coexistence, such fitness differences drive competitive exclusion and can be approximated by species average growth rate differences (Chesson 2000). In general, the species with the highest average fitness displaces all competitors in the absence of niche differences.

Despite the achievements observed in past years about the role of niche and fitness differences in community dynamics (MacDougall et al. 2009), species abundances has been coined as relevant mostly regarding niche perspectives (Adler et al. 2007), and when assuming neutral process, its role remains unknown in some degree (McGill et al. 2006). However, assuming the basic assumptions of neutral process formulations we can expect population abundance to play an important role on the outcome of fitness inequalities and consequently on community dynamics. For example, fitness inequalities can lead one population to dominate the community after a specific period (Adler et al. 2007). However, more individuals to "trade" in a finite and resource-conserved system can balance the chances of success that govern demographic properties, if these demographic properties are also influenced by stochasticity. Therefore, population initial abundance could work as an equalizing, by reducing the fitness inequalities, rather than stabilizing mechanism, promoting coexistence.

In this study, we aim to provide a theoretical evidence that, assuming the stochastic nature of neutral process on population demography, population initial abundance can work as an equalizing mechanism, diminishing fitness inequalities among species and influencing their rates of coexistence, here defined as persistence in time.

**Methods:**

*A Minimum Stochastic Competition Model:*

To test my predictions I used a simple probabilistic model with stochastic demography. This model can be visualized as an alternative interpretation for the formulas that would apply to classic neutral models. The community assembly follows a random demographic process in a hypothetical two species, A and B, system. Each species have specific chances to lose one individual or to reproduce, gaining one individual, at each time step. Once species are at a site, the abundance (abundance) of each will be proportional to the total amount of individuals at that point in time. The amount of individuals in the system is conserved, representing a closed system, and thus it is always constant and equal to the system carrying capacity *K*, a zero-sum neutral assumption (Hubbell 2001). When one individual dies in the community, it is replaced by a new individual of the competing population, simulating a lottery model among competing populations. This assumption strongly simplifies the model structure by directly relating reproduction and death probabilities, and guarantees that every time step the community will alter its proportion of individuals among populations, otherwise populations could be just replacing its own individuals keeping their relative densities constant. This situation could describe an exploitative competition scenario between two species where the amount of resources is finite and allocated in the biomass of the organisms inhabiting this system. When one individual dies the newly available resources is used by the other species to build a new individual. Therefore, we can simplify the probability of success of species A over species B, or vise-versa, as it chances to survive and reproduce with a single probability parameter, representing the fitness differences among species.

Let's assume two situations, first, a strict equivalence of species, in such a way that each species is equally likely to survive and reproduce; and second, species differs in their chances of success (i.e., to survive and reproduce). The probability of one individual of species A to survive and reproduce every time step can be represented by *P(A) = r* and the probability of B to survive and reproduce is *P(B) = q = 1 –r*. According to the classical Neutral Theory (Hubbell 2001) when *r* and *q* have the same probabilities (i.e, 1/2) one species will go extinct and the other species will dominate after a specific time interval, a probabilistic process known as random walk. However, given the same probabilities *r* and *q*, what are the chances of A outcompete

B and B goes extinct or the chances of B outcompete A and A goes extinct? To solve this question we must determine the initial population abundance of each species. The initial population abundance for species A is given by *n* and the initial population abundance for species B is given by *K-n*, where *K* is the total amount of individuals supported by the system (i.e., the carrying capacity). Thus, we can represent the chances of A to outcompete B given that A has *n* individuals as:

$$Pn = P(\text{A outcompete B} \mid \text{A has n individuals}) \quad (\text{eq. 3})$$

According to the law of total probability, we can re-write the current probability described in equation 3 as a recursive equation averaging the probabilities of the two possible next states, gaining one individual with probability *r* or losing one individual with probability *q*:

$$Pn = rP(n+1) + qP(n-1) \quad for \quad 1 \leq n \leq K-1 \ (\text{Eq. 4})$$

When *r* is not equal to *q*, the explicit solution for this recursive equation can be given by (A.1):

$$Pn = \frac{1-\left(\frac{q}{r}\right)^n}{1-\left(\frac{q}{r}\right)^K} \quad (\text{Eq. 5})$$

However in a classical neutral model, *r* and *q* must be equal, so that species are equivalent. When *r* is equal to *q* the quotient (*q/r*) equals to one. Therefore, assuming that *x = (q/r)* and substituting *x* in equation 5 we have:

$$Pn = \frac{1-x^n}{1-x^K} \quad \text{(Eq. 6)}$$

Solving the $\lim_{x \to 1} \frac{1-x^n}{1-x^K}$ we use the L'Hospital's rule and take the derivatives of both numerator and denominator to find the solution:

$$Pn = \lim_{x \to 1} \frac{n-x^{n-1}}{K-x^{K-1}} = \frac{n}{K} \quad \text{for } r=q \quad \text{(Eq. 7)}$$

Fitness inequalities can be acknowledged in the present model when *r* and *q* are not equal, in this case, species are not equivalent and one species has greater chances of surviving and reproducing over the other, which can be depicted by its abilities or specific traits (Shipley 2008). Thus, we can use equation 5 to represent a stochastic competition model between a pair of species. Equation 6 describes the outcome of this interaction when species are equivalent, in fitness terms, and shows that the probability of species A to outcompete B is directly proportional to its population abundance and inversely proportional to the total number of individuals in the system.

We can also consider the number of time steps that species are expected to coexist before either one goes extinct or dominates the community. Using the same rationale used for the probabilities of outcompeting the competing species, the expected number of time steps for a species with *n* initial individuals equals 1 plus the weighted average of the expectations for the two possible next stages, *n+1* or *n-1*. The function giving the expected number of time steps, assuming that species A have *n* individuals as its initial abundance, satisfies the following recursive equation:

$$Tn = rT(n + 1) + qT(n - 1) + 1 \quad \text{(Eq. 8)}$$

The solution of this recursive equation is (A.2):

$$Tn = \frac{\left(\frac{q}{r}\right)+1}{\left(\frac{q}{r}\right)-1}\left[n - K\left(\frac{1-\left(\frac{q}{r}\right)^n}{1-\left(\frac{q}{r}\right)^K}\right)\right] \textit{ for } r \neq q \quad \text{(Eq. 9)}$$

When *r* is equal to *q* the solution for eq. 8 can be achieved by taking the limit of the ratio *q/r* tending to one in equation 9, the solution for this limit is given by:

$$Tn = \left(\frac{K}{2}\right)^2 - \left(n - \frac{K}{2}\right)^2 \textit{ for } r=q \quad \text{(Eq. 10)}$$

In the model simulation, each time step corresponds to the time interval that one organism dies and another is added in the community. It is reasonable to think that this time interval is different for each pairwise organism interaction, due to differences in the life cycles of each species. Therefore, the unit of time used in this study is arbitrary and could be represented by different units.

**Results:**

Model results showed that assuming fitness equivalences (i.e, *r=q*) the probabilities of one species to dominate the community is directly proportional to its initial abundance. In Figure 1 we can observe that the probability of species A to outcompete species B follows a straight line for 0 to 1 across different initial population densities when the system support capacity (K) is equal to 20. It is also possible to observe that for fixed values of initial abundance the probability of species

A to dominate the community is reduced as K increases, in other words, when species B increases its fraction of the total number of individuals in the community.

When *r* and *q* are not equal, and so species A and B have different chances of success in the environment, the probabilities of one species to outcompete the other drastically changes (Figure 2). When species A have higher chances of success than species B (i.e., $q/r < 1$) the chances of A outcompeting B rapid increases with the increase in initial population abundance. However, when $q/r$ is greater than 1 the chances of A outcompeting B are much smaller, and are greater then 50% only when species A starts with more than half of the system carrying capacity. This result highlight the interactive effects between fitness inequalities and initial population abundances.

The model results also shows that, assuming the same initial densities among populations, as K increases the chances of species A to go extinct decreases if it hashigher chances to survive and reproduce than species B (i.e., $q/r < 1$), but increases if its chances are lower than B (Figure 3). This result is particularly interesting because it shows that fitness inequalities can be more important in competitive situations especially when total population abundance is high.

From equations 8 and 9 it is possible to observe that the expect time of coexistence among species, the time period necessary to one species dominate the community, be proportional to system support capacity. In Figure 5 we can also observe that the expected time of coexistence between two competing species is dependent on species initial densities and their chances of success. The expected time of species coexistence is lower if one species has low success and starts with lower abundance, and high if that species have low chances of success but starts with high abundance. For a fixed K, the maximum possible expected time of coexistence happens when both species have equal initial densities and have equal chances of success in the environment. On the other hand, when the q/r ratio is biased towards a species, unstable coexistence is diminished.

**Discussion:**

The relationship between population abundances and coexistence can be influenced by the covariance between competition and the environment (Chesson 2000), allowing populations to bounce back when rare, and has been coined as one of the main mechanisms behind species stable coexistence (Fox 2012). The present simulation does not account for this mechanism, instead it is centered in examine the role of neutral process and how neutral assumptions can justify population abundance as an equalizing rather than stabilizing mechanism, as in previous niche based models. The results provided by the present model shows that population abundance strongly determines the chances that one species has to outcompete other species and can alter the persistence in time of competing species. Traditional Lotka-Volterra competition models, and many other derived from it (Chesson 2000), also supports the idea that population abundance plays a major role in determining coexistence, or the dominance of a particular species. However, the mechanism behind the role of species abundance in these models is in essence stabilizing, since it influences the interplay between intra and interspecific competition (Adler et al. 2007). The role of species abundance as equalizing mechanism relies on the fact that fitness inequalities leads to differential survivorship, and having more individuals to "trade" in a finite and resource conserved system can counter-balance the chances of success that govern demographic properties. Although the present model lacks in its ability to provide a mechanistic framework linking the chances of success of a species and organism traits (Shipley et al. 206), it provides insightful hypothesis about the role of neutral assumptions on community dynamics in terms of unstable coexistence.

The model results show that under maximum fitness similarity, which is depicted by equal chances of the individuals from each population to survive and reproduce, the chances of one species to outcompete other are directly proportional to its abundance. These results are in accordance with the classical Neutral Theory and represent the outcome of the stochastic nature of population demography, resulting in a random walk that favors the population with more individuals (Hubell 2001). When considering the persistence in time of the populations under maximum fitness similarities, the model reveals that the more similar two species are in their chances of surviving and reproducing, the more likely that they persist together for long periods, especially when they present similar abundances, apparently due to fitness equivalency. Fitness similarities can in fact generate coexistence from weak stabilizing mechanisms (Cadotte 2007). This study also highlights that considering the stochasticity of demographic properties whether persistence in time is due to equalizing effects, they are conditional to total species abundance. This fact is corroborated if we consider the system support capacity (i.e., K) tending to infinity, which will result in an infinite period of species coexistence, when each species has a fixed fraction of individuals of K, which is irrespective to fitness equivalences.

Although previous studies have conducted neutral models simulations by evaluating the dynamics of just one species at time (McKane et al. 2000), in fact the abundances of different species are not independent of each other (Etienne and Olff 2005), especially if a zero- sum assumption is made. McGill et al. (2006) pointed out that it is unclear the effect of non-independence of species in the real world, although it clearly becomes bigger as the number of species and individuals becomes smaller.

The present model shows that when two populations present fitness inequalities, the chances of one species to outcompete other is strongly dependent on population abundance. Is this case, population abundance can work as an equalizing effect, counterbalancing fitness inequalities, when the less favorable species holds a higher fraction of total population abundance. In other words, population abundance can even make species with a much lower fitness be more likely to dominate the community as its abundance increases. On the other hand, fitness inequalities can be intensified when the species with higher chances of success also holds a higher fraction of the individuals in the community. Inherently, fitness inequalities prevails especially when system support capacity is high. The reason is that the bigger the system the further the population has to go before extinction. The population with a disadvantage can only win if it does so quickly (small numbers), large carrying capacities make it very hard to win. This result clearly shows the contrast between the role of relative and total population abundances in zero-sum stochastic competition scenarios. A biased relative abundance can compensate for fitness inequalities in terms of the chances of a species to outcompete others. But, as previously stated, high total population densities, or system carrying capacity, makes fitness inequalities to prevail in the resulting ecological outcome.

Opposing to the traditional stabilizing coexistence provided by ecological tradeoffs (Chesson 2000), the interplay between fitness inequalities and population abundance on species coexistence in the present model is in its essence unstable. From Chesson's (2000) framework, equalizing coexistence is ultimately unstable since fitness equivalency results in intrinsic rates of population increase that matches zero. The present model shows that populations subjected to abundance-independent mortality will not be able to recover (i.e., has less chances to recover) when in the presence of a competitor with equivalent fitness but with greater population abundance.

Although, the model was not originally developed to deal with applied issues, the results presented in this simulation can have important implications in real world scenarios. For example, the notion that ecological invasion should refer to individual populations, together with the notion that determinants of invasion success may act at the level of populations, not species (Collauti and MacIsaac 2004), highlights the importance of the results obtained it the present study for species invasion of novel habitats. MacDougall et al. (2009) showed that fitness inequalities are important for the establishment of invaders species but dominance is mainly determined by niche differences. This study shows that native populations may be strongly benefited against being outcompeted by an invasive population based on its relative population abundance, even if it is considered an inferior competitor (i.e., have a lower fitness). In addition, if invasive populations have lower densities than native populations at the beginning of invasion stages, their success of establishment will be conditional to its chances to survive and reproduce, which highlights the fact that invasive species should have specific traits that ultimately grant then a higher fitness in the community (Caño et al. 2008). The divergence between the present results and the results provided by classical species coexistence models (Chesson 2000, MacDougall et al. 2009) is that coexistence in the present simulation is unstable (i.e., represents persistence in time) and niche process are absent, thus all results are based on neutral dynamics, neglecting the effects of stabilizing mechanisms. Despite these limitations, the present model shows that species relative abundances and total abundance can be important in determining the fate of community dynamics even without invoking niche mechanisms.

Predictions from neutral theory has had important impacts on the current ecological understanding of many patterns concerned with community ecology or biogeography (Leibold and McPeek 2006).  Although some assumptions of neutral models have raised criticism due to their ecological significance (Poulin 2004), they also have been supported when communities are observed in a broad temporal or spatial scale (Chave 2004, Resindell et al. 2012). The present results shows that under a neutral framework, stochasticity in demographic properties and conservation of individuals, populations relative initial densities plays a major role determining unstable coexistence among species, working as a equalizing mechanism. On the other hand, system carrying capacity, or the total population abundance, benefits fitness inequalities to prevail in competitive interactions. Therefore, the role of species relative abundances is not restricted to niche perspectives, instead it plays an important role in neutral process affecting the outcome of fitness differences in community dynamics and coexistence patterns. In addition, in a probabilistic world with the absence of niche mechanisms, species with higher fitness will not necessary outcompete less fitted species, especially if less fitted species have higher population abundance.

Even though natural systems are not fully neutral, it does not imply that the results of the present model cannot promote the progress in ecological understanding (Rosindell et al. 2012). We encourage empiricists to measure or estimate species coexistence, or how long species co-occur in time. Future studies and empirical observations should test if populations relative abundances can in fact favor less fitted species in unfavorable competitive scenarios or if system support capacity (a measure of the total number of individuals in the community), in fact aids fitness inequalities to prevail on species coexistence outcomes.

**Acknowledgments:**

The authors are grateful to A. Caliman for helpful suggestions on the manuscript. Guariento R.D. received a scholarship from the Brazilian council for scientific and technological development to support this research.

**Appendix 1**:

The Probability of species A to outcompete species B given that species A starts with n individuals can be described as:

$$Pn = P(A\ outcompete\ B\ |\ A\ starts\ with\ n\ individuals)$$

This equation can be re-written by averaging the two possible states for species A, gaining one individual with probability r and losing one individual with probability q:

$$Pn = rP(n+1) + qP(n-1) \quad for \quad 1 \leq n \leq K-1$$

where:
$$Pn(n=0) = 0$$
$$Pn(n=K) = 1$$

The equation above gives us a second order recursive equation whose characteristic polynomial can be represented by:

$$x^2 - \frac{1}{r}x + \frac{q}{r} = 0$$

The roots for this polynomial equation are:

$$x1 = 1$$

$$x2 = \frac{1-r}{r} = \frac{q}{r}$$

The general form of the recurrence solution is a linear combination of successive powers of these two characteristic roots:

$$Pn = A(1)^n + B\left(\frac{q}{r}\right)^n$$

Where A and B are constants to be determined by the two boundary conditions, Pn(n=0) = 0 and Pn(n=K) = 1. Inserting these values gives the conditions

$$0 = A + B \qquad 1 = A + B\left(\frac{q}{r}\right)^K$$

Therefore:

$$A = \frac{1}{1-\left(\frac{q}{r}\right)^K} \qquad B = \frac{-1}{1-\left(\frac{q}{r}\right)^K}$$

Substituting these terms in the equation above, we get the solution:

$$Pn = \frac{1 - \left(\frac{q}{r}\right)^n}{1 - \left(\frac{q}{r}\right)^K}$$

**Appendix 2**:

The recursive equation that determines the number of expected times steps before one species goes extinct and the other dominates is given by:

$$Tn = rT(n+1) + qT(n-1) + 1$$

This equation must be re-written in terms of successive differences and can be described by:

$$Sn = Tn - T(n-1)$$

$$S(n+1) = \frac{q}{r}Sn - \frac{q}{r} + 1$$

This equation represents a linear fractional recurrence and its solution is given by:

$$Sn = \left(\frac{\left(\frac{q}{r}\right)+1}{\left(\frac{q}{r}\right)-1}\right)\left((Sn(0)) - \frac{\left(\frac{q}{r}\right)+1}{\left(\frac{q}{r}\right)-1}\right)\left(\frac{q}{r}\right)^n$$

The particular solution for *Tn* is given by:

$$Tn = \left(\frac{\left(\frac{q}{r}\right)+1}{\left(\frac{q}{r}\right)-1}\right)n + \left((Sn(0)) - \frac{\left(\frac{q}{r}\right)+1}{\left(\frac{q}{r}\right)-1}\right)\sum_{j=1}^{n}\left(\frac{q}{r}\right)^{j}$$

$$= \left(\frac{\left(\frac{q}{r}\right)+1}{\left(\frac{q}{r}\right)-1}\right)n + \left((Sn(0)) - \frac{\left(\frac{q}{r}\right)+1}{\left(\frac{q}{r}\right)-1}\right)\left(\frac{q}{r}\right)\frac{\left(\frac{q}{r}\right)^{n}-1}{\left(\frac{q}{r}\right)-1}$$

The particular solution is of the form $T_n = C_1 + C_2 (q/r)^n + [((q/r)+1)/((q/r)-1)]n$ for constants $C_1$ and $C_2$, and the homogeneous solution is of the form $A + B(q/r)^n$ for arbitrary constants A and B. The sum of both solutions is:

$$Tn = A + B\left(\frac{q}{r}\right)^{n} + \left(\frac{\left(\frac{q}{r}\right)+1}{\left(\frac{q}{r}\right)-1}\right)n$$

Assuming that Tn(0)=0 and Tn(K) = 0, A and B can be replaced and the final solution is given by:

$$Tn = \frac{\left(\frac{q}{r}\right)+1}{\left(\frac{q}{r}\right)-1}\left[n - K\left(\frac{1-\left(\frac{q}{r}\right)^{n}}{1-\left(\frac{q}{r}\right)^{K}}\right)\right] \text{ for } r \neq q$$

When r is equal to q the quotient (q/r) equals to one. Therefore, assuming that x = (q/r) and substituting x in the equation above, we have:

$$Tn = \frac{x+1}{x-1}\left[n - K\left(\frac{1-x^{n}}{1-x^{K}}\right)\right]$$

Solving the $\lim_{x \to 1} \frac{x+1}{x-1}\left[n - K\left(\frac{1-x^n}{1-x^K}\right)\right]$ we get:

$$Tn = \left(\frac{K}{2}\right)^2 - \left(n - \left(\frac{K}{2}\right)^2\right) \; for \; r=q$$

Figure 1 - Probability of species A to outcompete species B (*Pn*) across a gradient of initial population densities for species A for different system support capacities (K). Fitness are equal among populations.

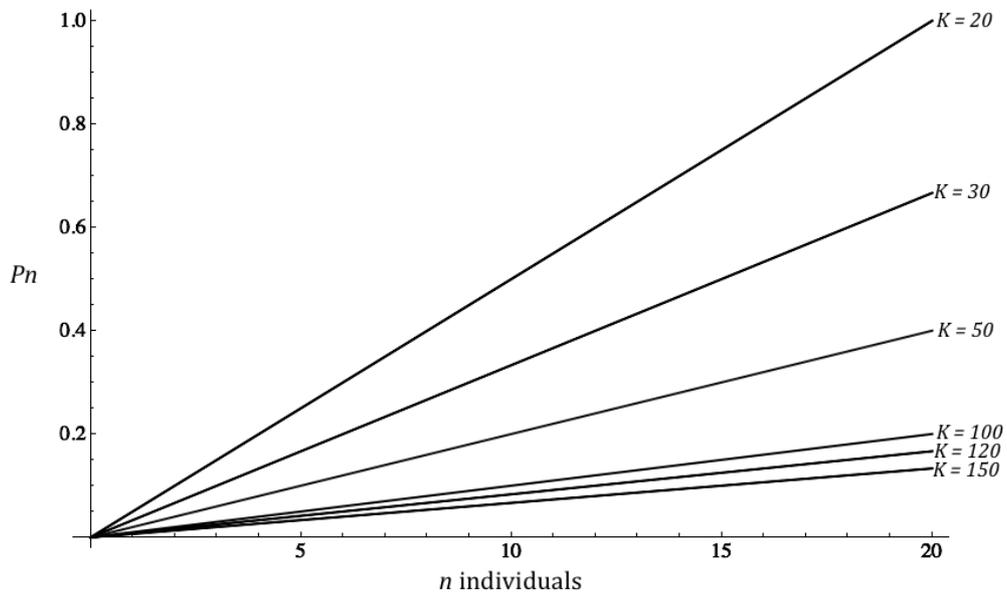

Figure 2 - Probability of species A to outcompete species B across a gradient of initial population densities for species A. The ratio *q/r* is defined as the ratio between the chances of one individual of species B to survive and reproduce over the chances of one individual of species A to survive and reproduce in the environment. System support capacity is set to K=20.

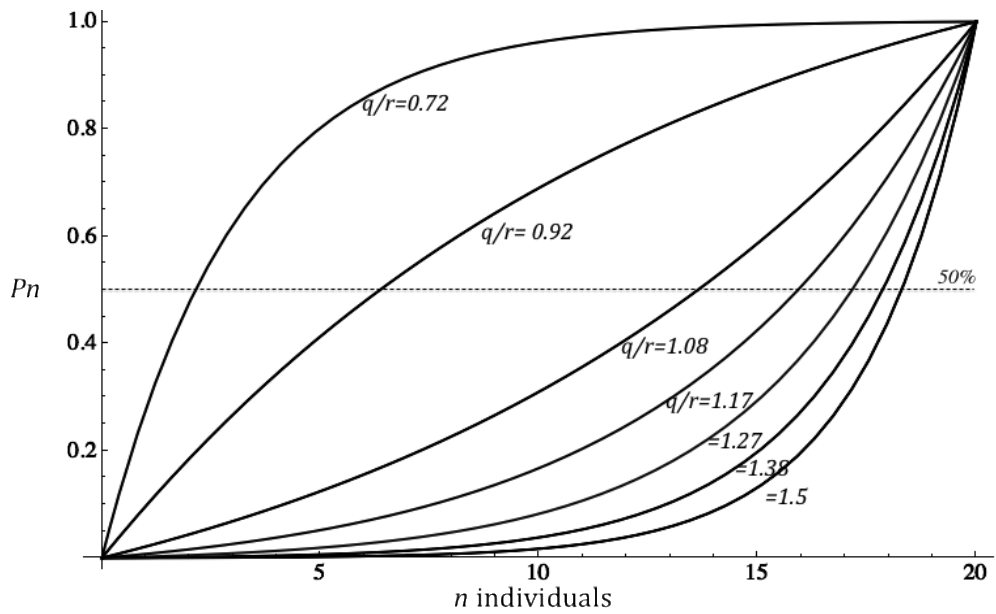

Figure 3 - Probability of species A to going extinct (Pe) across a gradient of system support capacity. The ratio *q/r* is defined as the same in figure 2. Both species share the same initial population abundances.

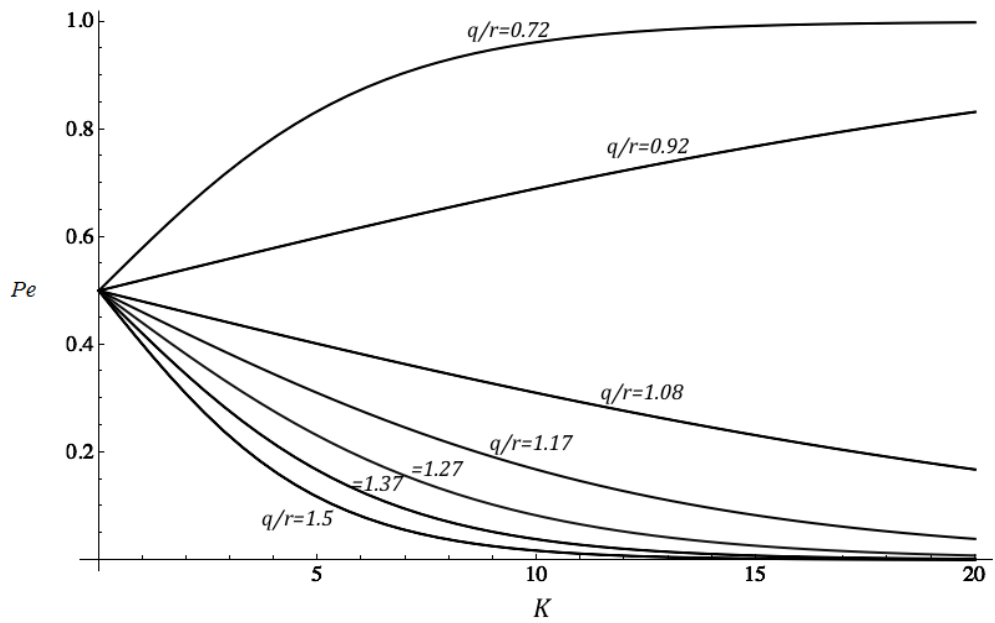

Figure 4 – Expected number of time steps (Tn) where both species A and B will coexist across a gradient of initial population densities for species A. The ratio *q/r* is defined as the same in figure 2. The system support capacity is K=20.

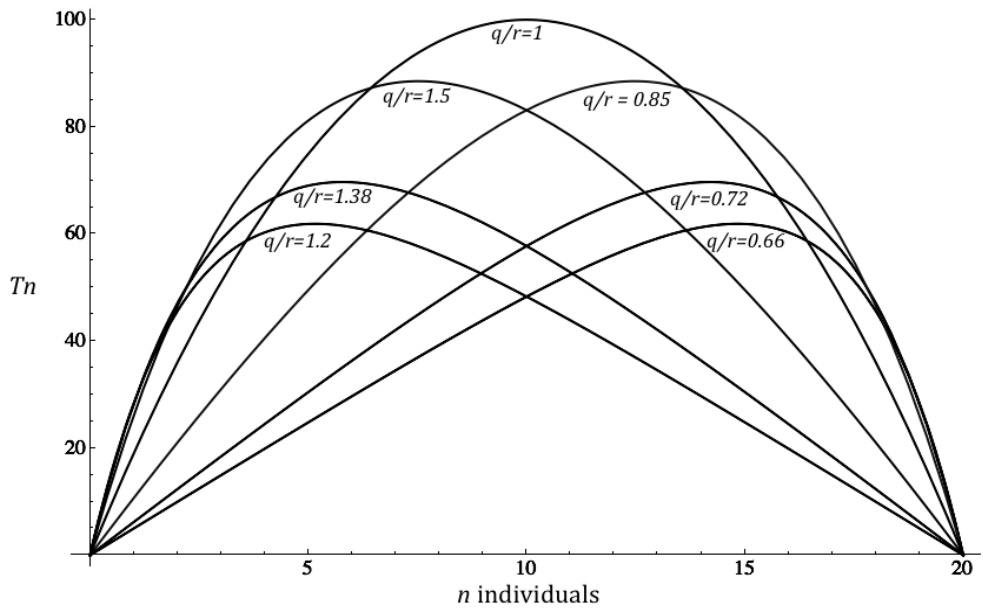